\documentclass[a4paper]{jpconf}
\usepackage{graphicx}
\usepackage{amsmath}
\usepackage{amssymb}
\usepackage{mathrsfs}

\usepackage{mathbbol}
\usepackage{bbding}

\usepackage{amsfonts}
\begin{document}
\title{The role of leptonic CPV phases in cLFV observables}

\author{J.~Kriewald$^{a}$\footnote{Speaker}, A.~Abada$^{b}$, A.~M.~Teixeira$^a$}

\address{$^{a}$ Laboratoire de Physique de Clermont (UMR 6533), CNRS/IN2P3,\\
Univ. Clermont Auvergne, 4 Av. Blaise Pascal, 63178 Aubi\`ere Cedex,
France

$^{b}$ P\^ole Th\'eorie, Laboratoire de Physique des 2 Infinis Irène Joliot Curie (UMR 9012), \\
CNRS/IN2P3,
15 Rue Georges Clemenceau, 91400 Orsay, France 
}

\ead{jonathan.kriewald@clermont.in2p3.fr}

\begin{abstract}
In models where the Standard Model is extended by Majorana fermions, interference effects due to the presence of
CP violating phases have been shown to play a crucial role in lepton number violating processes.
However, important effects can also arise in lepton number conserving, but charged lepton flavour violating (cLFV) transitions and decays.
Here we show that the presence of CP violating (Dirac and Majorana) phases can have a striking
impact for the predicted rates of cLFV observables.
We explore the interference effects in several cLFV observables, carrying for the first time a thorough analysis of the different observables and the implications for future observation.
We discuss how the presence of leptonic CP violating phases might lead to a loss of correlation between observables (typically present in simple SM extensions via heavy sterile fermions),
or even to the suppression of certain channels; these effects can be interpreted as suggestive of non-vanishing phases.
\end{abstract}

\section{Motivation}
In addition to constituting the first laboratory discovery of New Physics (NP), neutrino oscillations 
implied that neutral leptons are massive and that (neutral) lepton flavours are not conserved. In turn, this opens the door to new phenomena, forbidden in the Standard Model (SM), such as charged lepton flavour violation (cLFV) and leptonic CP violation (CPV).

Numerous SM extensions have been proposed to explain neutrino masses and leptonic mixings. Models in which right-handed neutrinos are added to the SM so that Dirac neutrino masses generated from the Higgs mechanism successfully accommodate oscilation data; however, these extensions are plagued by naturality issues (smallness of the Yukawa couplings, $Y^\nu$) and are very hard to test (for example, associated predictions for cLFV processes lying beyond any future experimental sensitivity).
Other (more appealing) possibilities include the different realisations of the seesaw mechanism. In particular, models  calling upon heavy Majorana neutral fermions (sterile states under the SM gauge group), as is the case of the type I seesaw~\cite{seesaw:I} and its low-scale variants variants (such as the inverse seesaw~\cite{Schechter:1980gr,Gronau:1984ct,Mohapatra:1986bd}, can be realised at low energies - close to the TeV -, leading to a very rich phenomenology, which encompasses cLFV and lepton number violation (LNV) processes. 

Numerous LNV processes (including neutrinoless double beta decays, or (semi)leptonic meson decays) are known to exhibit a strong dependence on leptonic CPV phases~\cite{Abada:2019bac}. In~\cite{Abada:2021zcm}, 
a thorough study of the effects of Dirac and Majorana phases on leptonic cLFV transitions and decays was carried, and in what follows we highlight the most relevant results.

\section{The role of phases: first approach}
We have considered an effective ``3+2 toy model'', in which 2 heavy neutral leptons (HNL) are added to the SM content. No assumption is made on the actual mechanism of neutrino mass generation - the spectrum contains 5 massive Majorana states, and leptonic  mixings are encoded in a $5\times5$ matrix, parametrised via 10 mixing angles $\theta_{\alpha j}$ and 10 CPV phases - 6 Dirac $\delta_{\alpha j}$ and 4 Majorana $\varphi_j$. In the limit of small mixing angles, the active-sterile mixings are given by 
\begin{equation}
\mathcal{U}_{\alpha (4,5)} \approx 
\left (\begin{array}{cc}
s_{14} e^{-i(\delta_{14}-\varphi_4)} &
s_{15} e^{-i(\delta_{15}-\varphi_5)} \\
s_{24} e^{-i(\delta_{24}-\varphi_4)} &
s_{25} e^{-i(\delta_{25}-\varphi_5)} \\
s_{34} e^{-i(\delta_{34}-\varphi_4)} &
s_{35} e^{-i(\delta_{35}-\varphi_5)} 
\end{array}
\right)\,,
\end{equation}
with $s_{\alpha i} = \sin \theta_{\alpha i}$
Notice that the would-be PMNS matrix is no longer unitary, which leads to modified charged and neutral lepton currents, and hence to significant contributions to several SM-forbidden observables.

\begin{figure}
    \centering
\mbox{\hspace*{-5mm}    \includegraphics[width=0.51\textwidth]{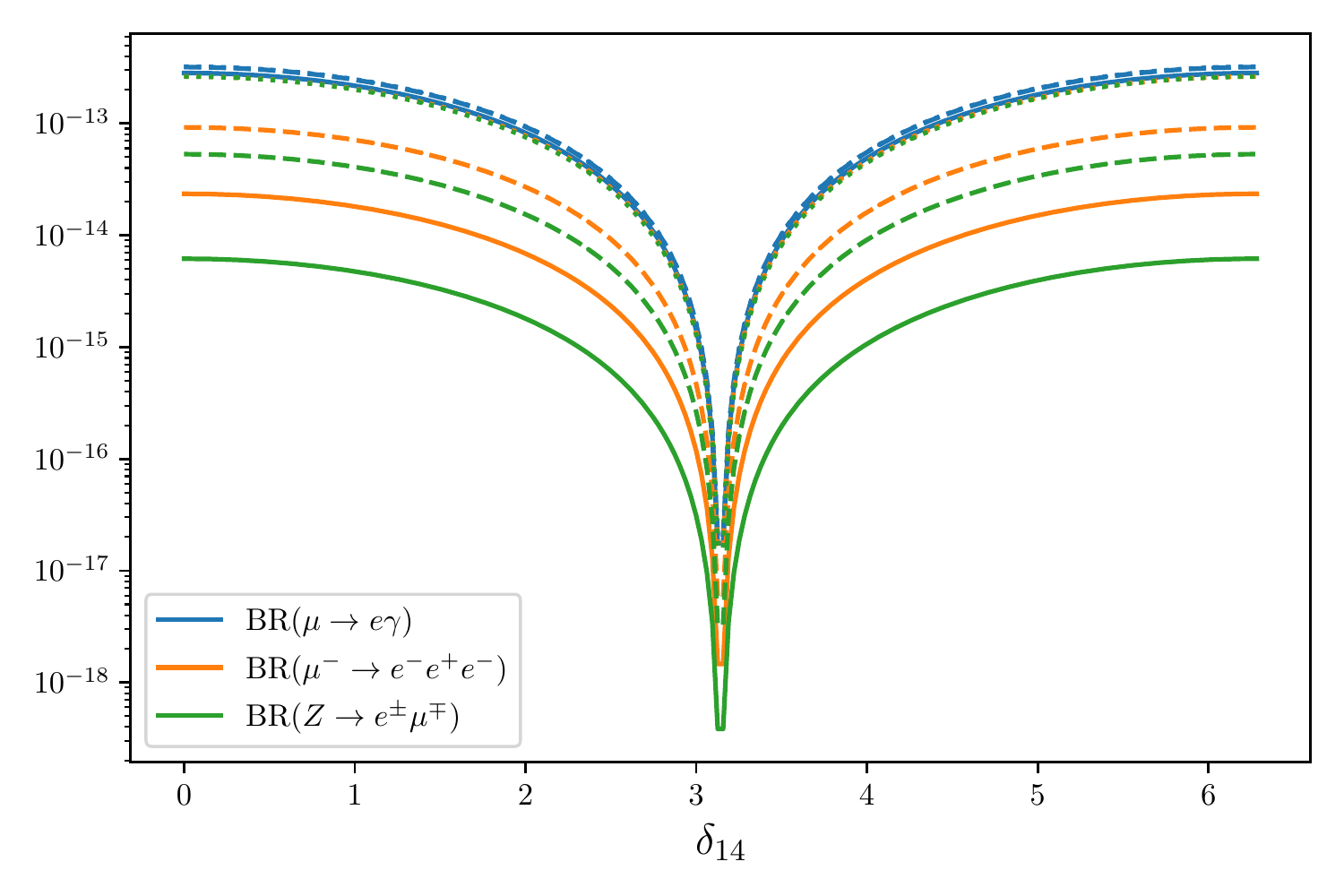}\hspace*{2mm}
\includegraphics[width=0.51\textwidth]{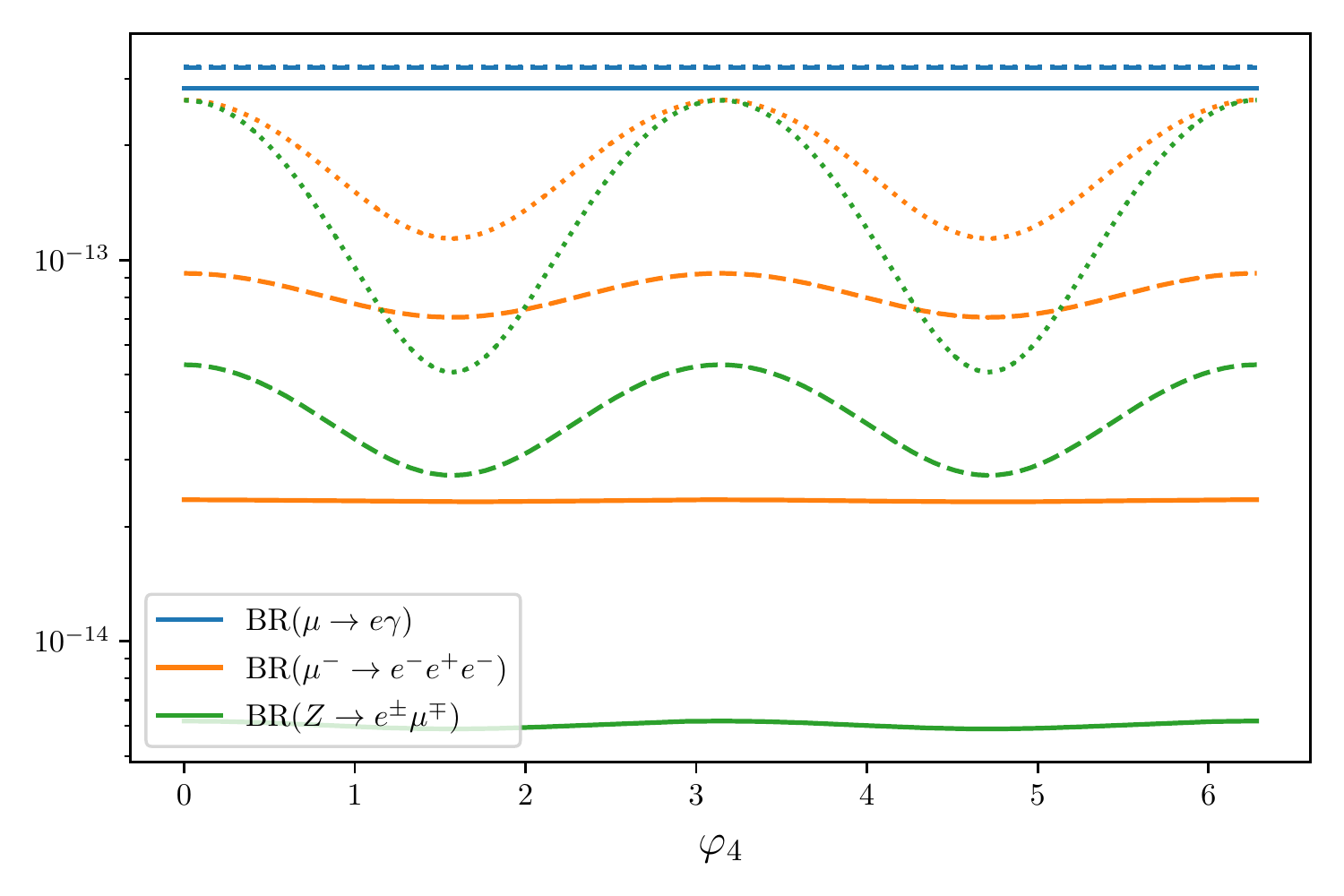}}
    \caption{ Dependence of cLFV observables on the CP violating Dirac phase $\delta_{14}$ (on the left) and Majorana phase $\varphi_4$ (on the right). Solid, dashed and dotted lines respectively correspond to $m_4=m_5=$ 1, 5, 10~TeV. From~\cite{Abada:2021zcm}.
        }
    \label{fig:1}
\end{figure}

In order to illustrate the role of CPV phases regarding cLFV observables, consider the case of $\mu \to e \gamma$ decays, mediated by $W$ bosons and both light and heavy neutrinos. 
The associated branching fraction (see~\cite{Abada:2021zcm}) is given by 
\begin{equation}
\text{BR}(\mu \to e \gamma)\propto 
       |G_\gamma^{\mu e}|^2\,, \text{with} \quad 
    G_\gamma^{\mu e} \, =\, \sum_{i=4,5} 
    \mathcal{U}_{e i}\, \mathcal{U}_{\mu i}^* \, 
    G_\gamma(m^2_{N_i}/M^2_W)\,.
    \end{equation}
In the limit $m_4 \approx m_5$ and for $\sin \theta_{\alpha  4} \approx \sin \theta_{\alpha  5} \ll 1$    
the form factor is given by
\begin{equation}
|G_\gamma^{\mu e}|^2 \approx 4 s_{14}^2 s_{24}^2 \cos^2\left(\frac{\delta_{14}+\delta_{25}-\delta_{15}-\delta_{24}}{2}\right)  G_\gamma^2(x_{4,5})\,.
\end{equation}
The cLFV rate clearly depends on the Dirac phases, with full cancellation obtained in the case $\delta_{14}+\delta_{25}-\delta_{15}-\delta_{24} = \pi$. 
Other form factors (for instance $Z$-penguins and boxes, relevant for three-body decays and muon-electron conversion, for example) also depend on the phases (both Dirac and Majorana phases), but have more involved associated expressions. 
The dependence of several $\mu-e$ cLFV observables on the Dirac phases  is shown on the left plot of Fig.~\ref{fig:1}, illustrated for $\delta_{14}$;  under the simple hypothesis 
$\sin \theta_{\alpha  4} =\sin \theta_{\alpha  5}$, and for $m_4=m_5=1$~TeV, one finds the above identified behaviour (and cancellation, for $\delta_{14}=\pi$), present for $\mu \to e \gamma$, $\mu \to 3 e$ and $Z\to e \mu$ decays. 
A similar dependence is found for Majorana phases in the considered observables (except for radiative decays, to which the Majorana CPV phases do not contribute). This is shown on the right panel of Fig.~\ref{fig:1}, for the same set of observables and underlying hypotheses.

Processes relying on different topologies (boxes, $Z$ and photon-penguins, ...) can exhibit a significant degree of interference (destructive or constructive) from the distinct contributions, so that Dirac and Majorana CPV phases can lead to cancellations or enhancements  of the associated rates. 
It is also important to mention that whenever $Z\nu\nu$ vertices are present, all flavours (and hence all phases) contribute.

\section{Towards realistic scenarios}
Following the above mentioned first simple approach, we now carry out a realistic study of the impact of CPV phases on cLFV observables; comprehensive scans of the parameter space are conducted (both for the mixing angles and all phases), and all available (relevant) constraints are applied. Concerning the latter, and in addition to the several cLFV constraints, we take into account experimental results and limits on SM extensions via TeV-scale HNL\footnote{We consider constraints from electroweak precision observables ($M_W$, $G_F$, invisible $Z$ width, ...), lepton universality tests (leptonic $W$ and $Z$ decays, ratios of leptonic meson decays, ratios of (semi)leptonic tau decays, ...), neutrinoless double beta decays, and finally perturbative unitarity constraints ($\Gamma_{N_{4,5}}/m_{{4,5}} \leq 1/2$); for a detailed description and corresponding references, see~\cite{Abada:2021zcm}.}.

\begin{figure}[ht!]
    \centering
\mbox{\hspace*{-5mm}    \includegraphics[width=0.51\textwidth]{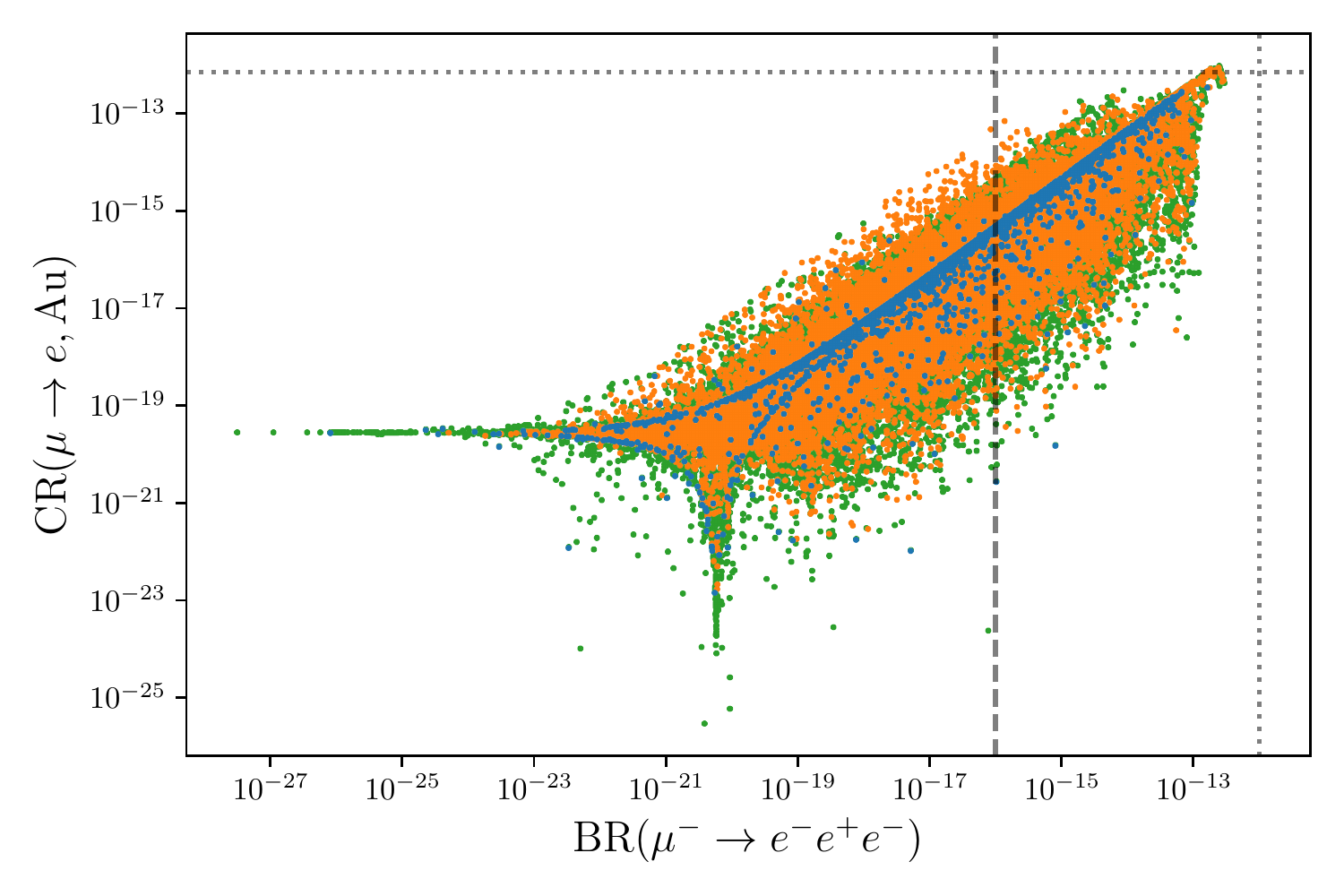}\hspace*{2mm}
\includegraphics[width=0.51\textwidth]{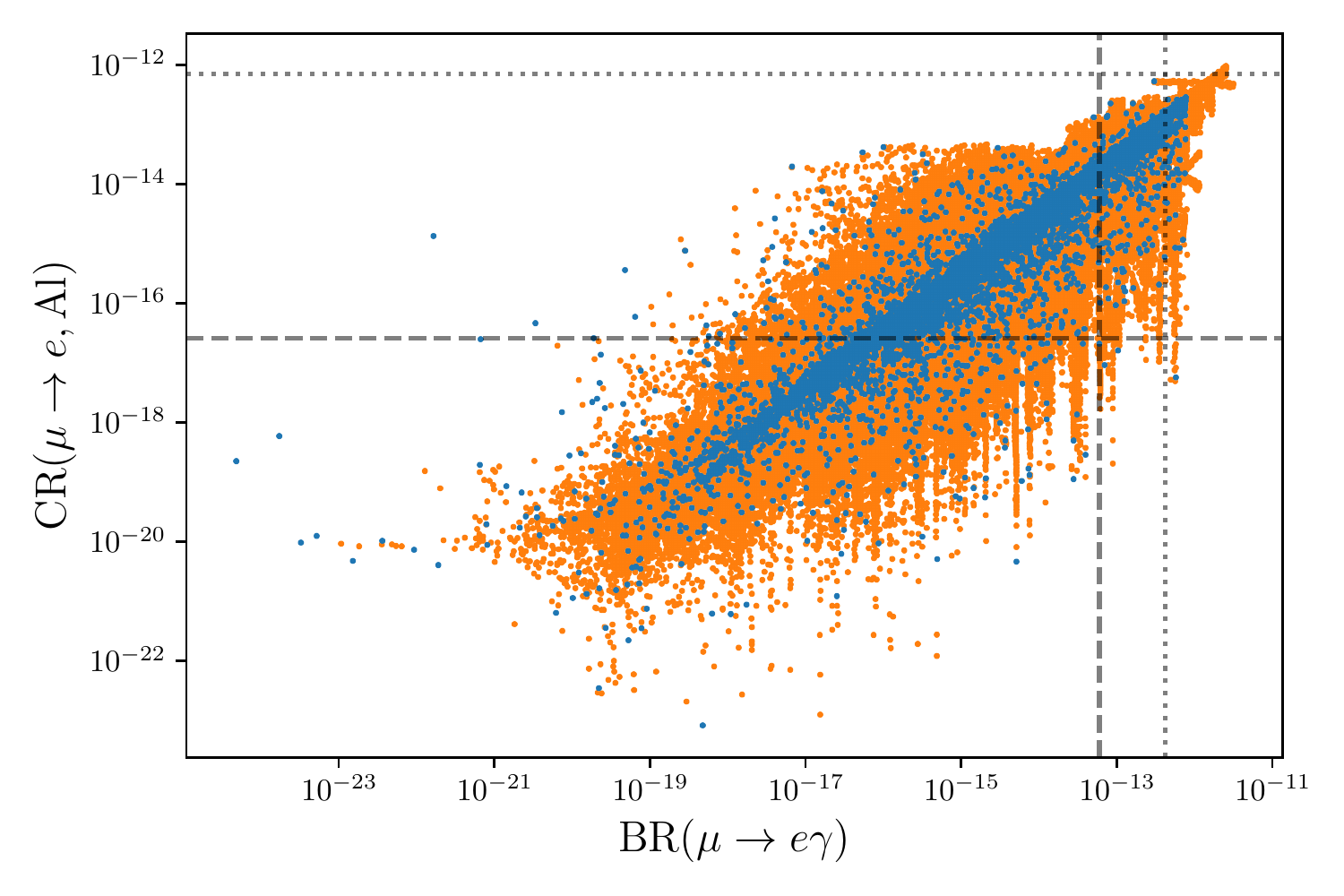}}
\caption{
    Correlation of $\mu-e$ cLFV observables, for varying values of the CPV Dirac and Majorana phases: vanishing values (blue), non-vanishing (orange), ``special grid'' (green), cf. description in text. From~\cite{Abada:2021zcm}.
        }
    \label{fig:2}
\end{figure}

On the left plot of Fig.~\ref{fig:2}, we display the effects of the CPV phases on the correlation between the rates of two $\mu-e$ sector observables, CR($\mu-e$, N) and BR($\mu \to 3e$). Leading to the results display, a random scan was performed over a semi-constrained parameter space: in particular, one now only imposes $\theta_{\alpha 4} \approx \pm \theta_{\alpha 5}$. We have taken degenerate heavy states ($m_4=m_5=1$~TeV), and for each point the CPV phases $\delta_{\alpha 4}$ and $\varphi_4$ were set to zero (blue points), randomly varied (orange) and further varied on a grid (green), the latter possibility aiming at ensuring that the special ``cancellation'' cases are included.
Since in the present mass regime for the heavy fermions both observables receive dominant contributions from $Z$-penguins, one expects that the associated rates be correlated; such a behaviour is indeed observed - cf. thick blue line of the CR($\mu-e$, N) vs. BR($\mu \to 3e$) plot. However, and once CPV phases are non-zero, one observes a loss of correlation, all the most striking for the ``special'' values of the phases $\{0, \frac{\pi}{4}, \frac{\pi}{2}, \frac{3\pi}{4}, \pi\}$ -  corresponding to the green points.   
In view of this behaviour, it is important to emphasise that HNL extensions of the SM should not be disfavoured upon observation of a single  cLFV signal; for example, 
should future collider searches strongly hint for the presence of sterile states with masses close to 1~TeV, and should  BR($\mu \to 3 e$)$\approx 10^{-15}$ be measured, one need not expect the observation of CR($\mu-e$, Al). While for vanishing CP phases the latter would be expected to be~$\approx \mathcal{O}(10^{-14})$, in the presence of CP violating 
phases, the expected range for the muon-electron conversion is vast, with CR($\mu-e$, Al) potentially as low as 
$10^{-18}$.

Finally, the results of a general overview of the ``3+2 toy model'' parameter space are displayed on the right panel of 
Fig.~\ref{fig:2}, in which we present 
CR($\mu-e$, N) vs. BR($\mu \to e \gamma$). The results stem from a comprehensive scan of the mixing parameters (all angles $\theta_{\alpha 4}$ and $\theta_{\alpha 5}$ independently varied) with all Dirac and Majorana phases 
randomly varied\footnote{The heavy states are no longer degenerate, but their masses are taken sufficiently close to allow for sizeable interference effects ($m_4=1$~TeV, with $m_5-m_4 \sim \Gamma_{N_{4,5}} \in [40~\text{Mev}, 210~\text{GeV}]$).}.
Again, one observes a dramatic loss of correlation (which would be otherwise present) for non-vanishing CPV phases; moreover, and as aforementioned, one can now have sizeable rates just for one of the observables. The experimental observation of $\mu \to e \gamma$ need not be accompanied by the observation of $\mu-e$ conversion in Aluminium (and vice-versa).

\section{Summary and conclusions}
As discussed here, the presence of CPV Dirac and/or Majorana phases can have a strong impact on the rates of cLFV observables, leading to a suppression or enhancement of the latter. 

The possible presence of the leptonic phases - which are a generic feature of mechanisms of neutrino mass generation - should be also taken into account upon interpretation of future data. CPV phases play a crucial role in the assessment of  viability of (regimes of) SM extensions via HNL. Several examples are provided in Table~\ref{table:1}, in which we summarise the predictions of benchmark points P$_i$ (for distinct choices for the active-sterile mixing angles) regarding cLFV observables, as well as the predictions associated with non-vanishing values of the phases (P$^\prime_i$):
\begin{eqnarray}
\label{eq:Pi:angles}
& \text{P}_1: &
s_{14} = 0.0023\,, \:s_{15} = -0.0024\,,\:
s_{24} = 0.0035\,, \:s_{25} = 0.0037\,, \:
s_{34} = 0.0670\,, \:s_{35} = -0.0654\,, \nonumber \\
& \text{P}_2: &
s_{14} = 0.0006\,, \: s_{15} = -0.0006\,, \: 
s_{24} = 0.008\,, \: s_{25} = 0.008\,, \: 
s_{34} = 0.038\,, \: s_{35} = 0.038\,, \nonumber \\
& \text{P}_3: &
s_{14} = 0.003\,, \: s_{15} = 0.003\,, \:  
s_{24} = 0.023\,, \:  s_{25} = 0.023\,, \:  
s_{34} = 0.068\,, \:  s_{35} = 0.068\,. 
\end{eqnarray}
The variants P$^\prime_i$ have identical mixing angles, but in association with the following phase configurations:
\begin{eqnarray}\label{eq:Pi:phases}
\text{P}^\prime_1:    
\delta_{14} = \frac{\pi}{2}\,, \:  
\varphi_4 = \frac{3\pi}{4}\,;\quad  
\text{P}^\prime_2:    
\delta_{24}=\frac{3\pi}{4}\,, \:  
\delta_{34} = \frac{\pi}{2}\,, \:  
\varphi_4 = \frac{\pi}{\sqrt{8}}\,; \quad  
\text{P}^\prime_3: 
\delta_{14}\approx \pi\,, \:  
\varphi_4\approx \frac{\pi}{2}\,.
\end{eqnarray}
\noindent We have chosen $m_4=m_5=5$ TeV for all three benchmark points.
As an example, notice that the regime of large mixing angles associated with P$_3$ would be excluded due to conflict with current limits; however, the presence of CPV phases allows to readily reconcile the predictions with observation (P$_3^\prime$), and thus to render viable the associated mixing regime.

In summary, the presence of leptonic CPV phases (both Dirac and Majorana) should be consistently included in phenomenological analysis of the prospects of HNL extensions of the SM in what concerns cLFV.

\section*{Acknowledgements}
JK thanks the Organisers of TAUP2021 for the invitation to present this work. We are grateful for the support from the European Union's Horizon 2020 research and innovation programme under the Marie Sk\l{}odowska-Curie grant agreement No.~860881 (HIDDe$\nu$ network) and from the IN2P3 (CNRS) Master Project, ``Flavour probes: lepton sector and beyond'' (16-PH-169). 

\renewcommand{\arraystretch}{1.}
\begin{table}[h!]
\caption{Predictions for cLFV observables in association with P$_i$, and variants with non-vanishing CP violating phases, P$_i'$. The symbols 
({\small\XSolidBrush}, $\checkmark$, $\circ$)  denote 
rates  in conflict with current experimental bounds, predictions within future sensitivity and those beyond future reach.
}
\vspace*{2mm} \hspace*{10mm}
    \begin{tabular}{|l|c|c|c|c|c|}
    \hline
 & BR($\mu \to e\gamma$) & BR($\mu \to 3e$) & CR($\mu - e$, Al) & BR($\tau \to 3\mu$)& BR($Z \to \mu \tau$)\\
 \hline\hline
$\text P_1$ & $ 3\times 10^{-16}$ \:\:$\circ$ & 
$ 1\times 10^{-15}$ \:\:$\checkmark$& 
$ 9\times 10^{-15}$ \:\:$\checkmark$& 
$ 2\times 10^{-13}$ \:\:$\circ$&
$ 3\times 10^{-12}$ \:\:$\circ$\\
$\text P_1^\prime$
& $ 1\times 10^{-13}$ \:\:$\checkmark$& 
$2\times 10^{-14}$ \:\:$\checkmark$& 
$1\times 10^{-16}$ \:\:$\checkmark$& 
$1\times 10^{-10}$ \:\:$\checkmark$& 
$2\times 10^{-9}$ \:\:$\checkmark$\\
\hline
\hline
$\text P_2$ 
& $2\times 10^{-23}$ \:\:$\circ$
& $2\times 10^{-20}$ \:\:$\circ$ 
& $2\times 10^{-19}$ \:\:$\circ$ 
&  $1\times 10^{-10}$ \:\:$\checkmark$
& $3\times 10^{-9}$ \:\:$\checkmark$\\
$\text P_2^\prime$
& $6\times 10^{-14}$ \:\:$\checkmark$
& $4\times 10^{-14}$ \:\:$\checkmark$
& $9\times 10^{-14}$ \:\:$\checkmark$
&  $8\times 10^{-11}$ \:\:$\checkmark$
& $1\times 10^{-9}$ \:\:$\checkmark$\\
 \hline
 \hline
 $\text{P}_3$ 
 & $2\times 10^{-11}$ \:\:{\footnotesize \XSolidBrush}
 & $3\times 10^{-10}$ \:\:{\footnotesize \XSolidBrush} 
 & $3\times 10^{-9}$ \:\:{\footnotesize \XSolidBrush}
 & $2\times 10^{-8}$ \:\:$\checkmark$
 & $8\times 10^{-7}$ \:\:$\checkmark$\\
$\text P_3^\prime$
& $8\times 10^{-15}$ \:\:$\circ$
  & $1\times 10^{-14}$ \:\:$\checkmark$
  & $6\times 10^{-14}$ \:\:$\checkmark$
  & $2\times 10^{-9}$ \:\:$\checkmark$
  & $1\times 10^{-8}$ \:\:$\checkmark$\\
 \hline
\end{tabular}\label{table:1}
\end{table}
\renewcommand{\arraystretch}{1.}

\end{document}